\documentclass{aa}  

\usepackage{graphicx}
\usepackage{txfonts}
\begin{document}

   \title{Rotationally resolved spectroscopy of dwarf planet (136472) Makemake}

   \author{V. Lorenzi
          \inst{1}
          \and
          N. Pinilla-Alonso\inst{2}
          \and J. Licandro \inst{3,4}
          }

   \institute{Fundaci\'on Galileo Galilei-INAF, Spain\\
              \email{lorenzi@tng.iac.es}
         \and Department of Earth and Planetary Sciences, University of Tennessee, TN, USA
         \and Instituto Astrof\'isico de Canarias, IAC, Spain
         \and Departamento de Astrof\'isica, Universidad de La Laguna, Tenerife, Spain
             }

   \date{Received September 15, 1996; accepted March 16, 1997}

 
  \abstract
   {Icy dwarf planets are key for studying the chemical and physical states of ices in the outer solar system. 
   The study of secular and rotational variations gives us hints of the processes that contribute to the evolution of their surface.}
   {The aim of this work is to search for rotational variability on the surface composition of the dwarf planet (136472) Makemake}
   {We observed Makemake in April 2008 with the medium-resolution spectrograph ISIS, at the William Herschel Telescope (La Palma, Spain) 
and obtained a set of spectra in the $0.28$ - $0.52\ \mu m$ and $0.70$ - $0.95\ \mu m$ ranges, covering 82\% of its rotational period. 
 For the rotational analysis, we organized the spectra in four different sets corresponding to different rotational phases, and after discarding one with low signal to noise,
 we analyzed three of them that cover $71\%$ of the surface. 
For these spectra we computed the spectral slope and compared the observed spectral bands of methane ice with reflectances of pure methane ice 
to search for shifts of the center of the bands, related to the presence of $CH_4/N_2$ solid solution.}
   {All the spectra have a red color with spectral slopes between 20\%/1000 \AA\ and 32\%/1000 \AA\, in accordance with previously reported values. 
Some variation in the spectral slope is detected, pointing to the possibility of a variation in the surface content or the particle size of the solid organic compound. 
The absorption bands of methane ice present a shift toward shorter wavelengths, indicating that methane (at least partially) is in solid solution with nitrogen. 
There is no variation within the errors of the shifts with the wavelength or with the depth of the bands, 
so there is no evidence of variation in the $CH_4/N_2$ mixing ratio with rotation. 
By comparing with all the available data in the literature, no secular compositional variations between 2005 and 2008 is found.}
   {}


   \maketitle

\section{Introduction}

The dwarf planet (136472) Makemake (hereafter Makemake) is a classical trans-Neptunian object (TNO) 
($a,e,i = 45.669\ AU,0.157, 29.01^{\circ}$, from the Minor Planet Center), with an estimated diameter $D\sim 1400\ km$ 
and an albedo $p_v=77\%$ \citep{Ortiz}. 

Icy dwarf planets are TNOs that are large and cold enough to retain large amounts of
volatiles on their surfaces \citep{schbrovol, levi2}. 
Three of the largest ones, Makemake,  (134340) Pluto (hereafter Pluto), and (136199) Eris (hereafter Eris), share spectral characteristics
with visible and near-infrared spectra dominated by methane-ice bands and with a red slope in the visible that 
suggests the presence of complex-organic materials on the surface  \citep{liceris,licmak, broeris,tegler2007}.
The presence of frozen methane on the surfaces of three large TNOs is an argument for the process suggested by \cite{spencer}, 
where surface methane is replenished from the interior, 
being ubiquitous in some icy dwarf planets. 
Makemake provides an exciting laboratory for studying processes also considered active
for Pluto: volatile mixing and transport, atmospheric freeze-out and escape, ice chemistry, and nitrogen phase transitions.

Some of Pluto's properties are closely related to these processes: (1) the presence of other volatiles like $N_2$ and $CO$ on its surface;,
(2) the presence of a global atmosphere composed  of $N_2$ and $CH_4$ \citep[e.g][]{Lellouch2009}, 
(3) the surface inhomogeneous composition with regions of different concentration of ices and some kind of complex organics, such as tholins.

Direct detection of  $N_2$ and $CO$ on the surface of Pluto was first done by means of near-infrared spectroscopy \citep{owen,quirico}. 
These volatiles are transparent in the visible region but can be detected by the presence of very weak absorption bands in the near-infrared. 
In Pluto’s spectroscopy, the hexagonal $\beta$  phase of $N_2$ ice is detected by means of its 2.15 $\mu m$ absorption band, 
and $CO$ ice is detected by means of a pair of narrow bands at 2.35 and 1.58 $\mu m$. 
Although  the 2.35 $\mu m$ band is at least one order of magnitude greater than that at 1.58 $\mu m$, 
it is also at the bottom of the 2.32 $\mu m$ absorption of $CH_4$ ice (that dominates the spectrum), so it is sometimes complicated to study.
The combination of temperature and diameter for Eris suggests that this object should retain a lot of volatiles similar to Pluto; 
however, no direct detection other than methane has been possible to date 
\citep[][and references therein]{Alvaro}. 
 The phase diagram of $N_2$ shows that at temperatures below 35.6 K, 
the phase of the $N_2$ ice changes from the $\beta$ phase to the $\alpha$ phase, and this affects the position, the depth, and the shape of the absorption bands. 
Experiments from \cite{quirico_b} show, for example, that the bands in the reflectance of $CO$ diluted in $\alpha$-$N_2$ are stronger but so narrow 
that they are far from the resolving power of the actual instrumentation. 
This fact is used by \cite{Alvaro} to explain why $N_2$ and $CO$ is not detectable on the surface of Eris, whose surface temperature is estimated below this limit of 35.6 K.

Considering Makemake's size ($\sim 1400$) and surface temperature ($\sim37\ K$), the retention regime for this object is different 
than the retention regime for Eris and Pluto \citep{brownrew}. 
Makemake is capable of retaining $CH_4$ but is not expected to retain large amounts of nitrogen or carbon monoxide ice, 
so its direct detection in its near-infrared spectrum is not probable \citep{brown2007}.
Alternatively, the presence of $N_2$ can be inferred in the visible by comparing the observed central wavelength of the $CH_4$ bands with laboratory data of pure $CH_4$ ice. 

Pluto's $CH_4$ bands are seen to be partially shifted (between $-12$ to $-20$ \AA) to shorter wavelengths relative to the wavelengths of pure methane-ice absorption bands, 
indicating that at least some of the methane ice on Pluto's surface is diluted in $N_2$ \citep{quirico, schmitt, doute}.
\cite{liceris} measured a shift of $-15 \pm 3$ \AA\ in the strong methane band around 0.89 $\mu m$ for Eris. Further work has confirmed the dilution of  $CH_4$ in  $N_2$ on the surface of this dwarf planet
because of the shift of several methane bands in the visible and near-infrared spectrum of this TNO 
\citep{dumas,merlin2009,aber,tegler2010,Alvaro,tegler2012}.
 \cite{licmak} and \cite{tegler2007, tegler2008} find that the centers of the absorption bands of methane ice 
in the visible and the near-infrared  spectra of Makemake's surface are shifted slightly blueward by $\sim -4$ \AA, 
suggesting that methane is in solid solution with nitrogen but with a  $CH_4/N_2$ fraction greater than in the case of Pluto and Eris. 
Recent studies support these results; in fact, using photometric observations of an stellar occultation, \cite{Ortiz}  concluded that Makemake has no global atmosphere like Pluto, 
and attribute this to a low abundance of $N_2$ ice.  
However, a local atmosphere could be possible \citep{Ortiz, stern2008}, in which case it could be related to the existence of patches at different temperatures on the surface of Makemake detected by thermic observations \citep{lim}. 
Surface regions of different concentrations of ices and organics with different albedos play an important role in the sublimation and recondensation of ices. 
Cyclical changes in the depths of $CH_4$ bands modulated by Pluto's diurnal rotation were found by \cite{grundy1996} and were used to constrain the longitudinal distributions of the three ice species on its surface.

In this paper we present visible spectra of Makemake obtained with the ISIS spectrograph attached to the 4.2 m William Herschel Telescope (WHT)  
at the Roque de Los Muchachos Observatory (ORM, La Palma, Spain) at different rotational phases covering $\sim$ 70\% of its rotation period,
with the aim of searching for rotational variability on the surface of Makemake. We also study the possibility of secular variations 
comparing our spectra with other spectra of Makemake from the literature. 
The paper is organized as follows. In Section \ref{obs_dat} we  describe the observations and data reduction; 
in Section \ref{res}  we analyze the obtained spectra; we compute the spectral slopes of the different phases and also of the other methane-ice rich dwarf planets; 
and we compute the shifts of the center of methane-ice absorption bands with respect to the center of pure methane-ice bands; 
finally, in Section \ref{dis}, we discuss the results, compare them with data in the literature, and present our conclusions.


\section{Observations and data} \label{obs_dat}

We observed Makemake in April 2008 with the 4.2 m William Herschel Telescope (WHT) 
at the Roque de Los Muchachos Observatory, La Palma Island, Spain,  
with the double armed, medium-resolution spectrograph ISIS, which enables simultaneous observation in both blue and red arms. 
We used a $1''$ slit, the R300B grating ($0.28$ - $0.52\ \mu m$, with a dispersion of 0.86 \AA/px) in the blue arm, 
and the R316R grating ($0.70$ - $0.95\ \mu m$, with a dispersion of 0.93 \AA/px) in the red one.

\begin{table}
         
          \begin{tabular}{llccc}
            \hline
            \hline
             \textbf{Date} &  \textbf{Time (U.T.)} & \textbf{Airmass} &  \textbf{Texp.}& \textbf{Phase} \\
             \hline
             2008/04/11 & 22.93 - 0.23 & 1.05 - 1.00 & 4500 s& phase 1 \\ 
             2008/04/12 & 0.72 - 2.27 & 1.01 - 1.11 & 5400 s &phase 2\\ 
             2008/04/12 & 2.67 - 4.22 & 1.17 - 1.60 & 5400 s &phase 3 \\ 
             2008/04/12 & 4.45 - 5.23 & 1.71 - 2.28 & 2700 s &phase 4\\ 
             \hline
          \end{tabular}
       \caption{Date, time (U.T.), and airmass of the observations and total exposure time corresponding 
to the four groups of Makemake spectra.}
\label{obs}
        
   \end{table}

We observed Makemake during a total of five hours on object  
and obtained 20 spectra of 900~s from both the blue and red gratings. 
Our observations covered 82\% of one rotation of Makemake \citep[$P_r$ = 7.771 $\pm$ 0.003 hours,][]{heinze,thirouin} over a time period of 6.3 hours.

We organized the spectra into four sets (see Table \ref{obs}).  
If we divided the rotational period of Makemake in four phases, starting arbitrarily at 2008/04/11 22.5 U.T., 
each group of spectra corresponds to data taken in a different phase.

Because the signal to noise ratio (S/N) of single raw spectra was too low to extract them, for each set of data 
we combined all the spectra to reach a reasonable S/N, 
and we extracted and calibrated the resulting spectrum following standard procedures using IRAF.
The last group of spectra (phase 4) had a poor S/N, and  
it was impossible to extract the spectrum from the blue grating, even after combining them. 
The spectrum from the red grating was extracted and calibrated, 
but its quality was too poor to make any kind of analysis, so we decided to discard this set of data. 
At the end, we obtained three combined spectra, each one with a blue part, covering the $0.35$ - $0.52\ \mu m$ range, and a red part, 
covering the $0.70$ - $0.92\ \mu m$ range and corresponding to three different rotational phases that cover the 70\% of the surface.

To correct spectra for telluric absorptions and to obtain the relative reflectance, we observed during the night
three stars that are usually used as solar analogs:
BS4486, Landolt (SA) 102-1081, and Landolt (SA) 107-684 \citep{land}. 
We observed them at different airmasses to cover the same range of airmass of the object.

\begin{figure}
   \begin{center}
   \includegraphics[width=9cm]{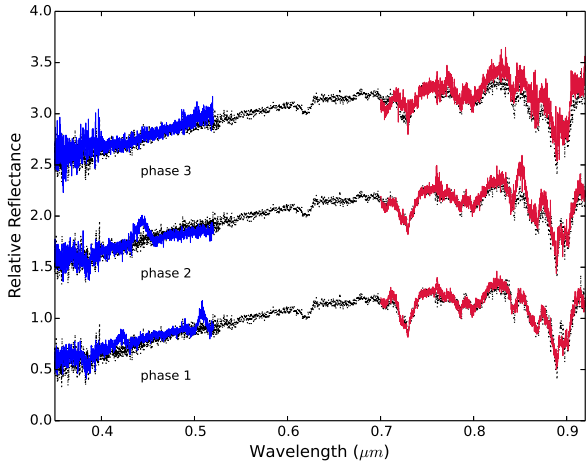}  
   \end{center}
      \caption{From bottom to top: the three final spectra (blue and red lines), shifted by 1 in relative reflectance for clarity, 
      superposed to  Makemake's spectrum from \cite{licmak} (black points) normalized at $0.55\ \mu m$.}
         \label{red_blue}
\end{figure}

Finally, each Makemake spectrum was divided separately by the spectrum of the solar analogs (at least two of them) 
that were observed at an airmass that was as similar as possible to the airmass of the object
and normalized at 0.43 $\mu m$ (blue part) and at 0.75 $\mu m$ (red part),  
thus we obtained the reflectance of Makemake relative to each solar analog.
Then, we averaged them for each rotational phase. 
These three final relative reflectances are shown in Fig. \ref{red_blue}. We finally used the spectrum from \cite{licmak}, 
normalized at $0.55\ \mu m$, to connect the blue and red parts of the spectra.

\section{Analysis and results} \label{res}

There are different parameters that will help us to extract information from our data:
(1) the slope of the relative reflectances, which  
indicates the presence of complex organics on the surface of the object;  and
(2) the center of the  methane ice absorption bands, 
which provides information about the mixing ratio of methane and nitrogen \citep{quirico}. 
For this purpose, we analyzed the blue and red parts of the spectra separately.

\subsection{Spectral slope}

The blue part of our set of spectra is featureless, apart from some structures 
that are probably observational or reduction artifacts, 
and presents a red slope.
To calculate the slope values, we normalized the spectra at $0.55\ \mu m$ 
and the fitted each spectrum between 0.40 $\mu m$ and 0.52 $\mu m$ with a linear function. 
The values obtained for the slope are shown in Table \ref{slope}, along with the associated error. 
The errors in the slope values are computed following the law of propagation of uncertainty, 
taking the error in the fit and the error introduced by dividing by the spectra of the solar analogs into account.

To evaluate the error introduced when dividing by the spectra of the solar analogs, 
we computed the slope of the division of each spectrum of the solar analogs observed during the night by one of them, 
following the same procedure as used to compute the relative reflectance of Makemake.
After comparing all the slopes, 
we found that the difference between the spectra of different solar analogs 
or between different spectra of the same standard at similar airmass is not bigger than 4\%/1000 \AA\  
(after discarding those that are not suitable owing
for example to probably incorrect centering in the slit). 
We took this value as systematic error. We also evaluated the error in the fit for each phase to calculate the slope, and
this error was added quadratically to the systematic error.
In Table \ref{slope} we also include the slopes of the spectra of other methane-ice rich dwarf planets computed by us following the same procedure as we used for Makemake. 

\begin{table}
         
          \begin{tabular}{lcc}
            \hline
            \hline
             \textbf{Data} &  \textbf{Slope (0.40-0.52 $\mu m$)} & \textbf{Error}\\
             \hline
             phase 1 & 23\%/1000 \AA & 4\%/1000 \AA  \\ 
             phase 2 & 20\%/1000 \AA & 4\%/1000 \AA  \\ 
             phase 3 & 32\%/1000 \AA & 4\%/1000 \AA  \\ 
             \cite{licmak} & 26\%/1000 \AA & 4\%/1000 \AA\\
             \hline
             \hline
             Eris$^a$  & 12\%/1000 \AA & 4\%/1000 \AA \\
             Pluto$^b$  &28\%/1000 \AA & 4\%/1000 \AA\\
             \hline
          \end{tabular}
       \caption{Top part of the table: slopes and respective errors computed for our three spectra 
       and for the spectrum of Makemake from \cite{licmak}. 
       Bottom part: slopes and respective errors calculated for the spectrum of Eris from (a) \cite{Alvaro} 
       and for the spectrum of Pluto from (b) Pinilla-Alonso N. (private communication). 
       All the slopes were computed between 0.40 - 0.52 $\mu m$ and for spectra normalized at $0.55\ \mu m$.
       }
\label{slope}
        
   \end{table}

\subsection{Shifts of the methane absorption bands}

\begin{figure}
   \begin{center}
   \includegraphics[width=9cm]{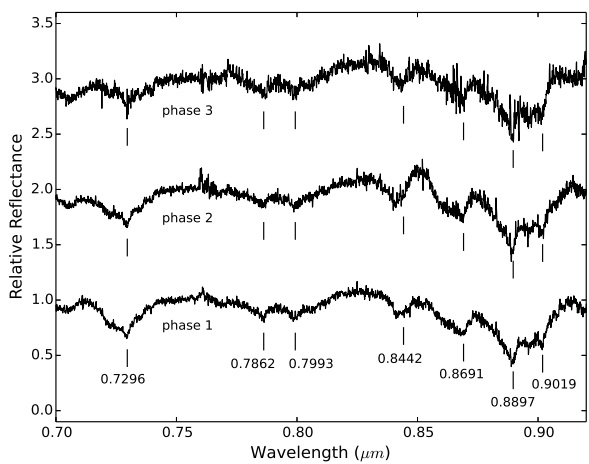}  
   \end{center}
      \caption{The red part of the three spectra corresponding, from bottom to top, to phases 1, 2, and 3, 
      normalized at $0.75\ \mu m$ and shifted by 1.0 in relative reflectance for clarity. 
      For the three spectra, the center of absorption bands of pure methane ice are marked.}
         \label{red_3}
\end{figure}

The red part of Makemake's spectrum from 0.70 $\mu m$ to 0.92 $\mu m$ is characterized by the presence of strong methane-ice absorption bands. 
In the three spectra we detected the $CH_{4}$ bands centered at 0.7296, 0.7862, 0.7993, 0.8442, 0.8691, 0.8897, and 0.9019 $\mu m$ 
(see Fig. \ref{red_3}). 
To search for possible shifts of the center of methane-ice absorption bands, 
we compared the absorption bands in our spectra with absorption bands in the relative reflectance of pure methane ice obtained
using a Shkuratov model \citep{shku} and the optical constants at 40 K from \cite{grund}.
The broadness of the absorption band present in the relative reflectance of an icy body depends inversely on 
the thickness of the ice layer that the light samples. This means that, on average, the weaker the band, the thicker the layer \citep[see ][]{liceris}.
Since the absorption bands of our spectra could be produced from light scattered by different layers, 
we decided to compare each band separately with a set of models corresponding to different grain sizes.

\begin{figure}
   \begin{center}
   \includegraphics[width=9cm]{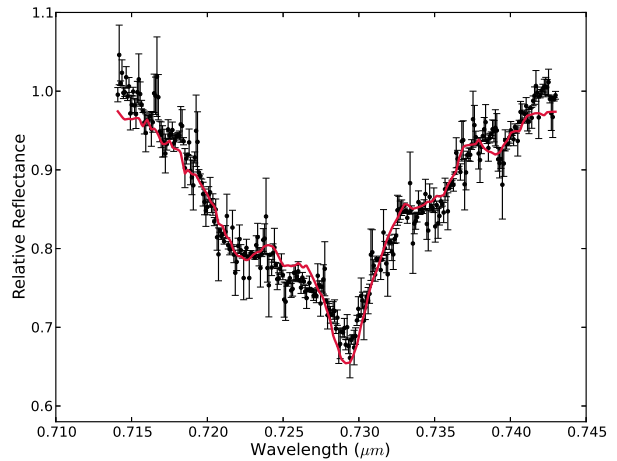}  
   \end{center}
      \caption{Example of the comparison of a model (red line) with the  $CH_4$ absorption band at $0.7296\ \mu m $ for the phase 1 (black points). 
      The center of the band was shifted by $-3.8$ \AA,\ and the model was obtained with a grain size of $ 4450\ \mu m $.}
         \label{fit}
\end{figure}

After dividing each spectrum by a continuum, obtained by fitting a spline function, 
we compared each band with various reflectances using a $\chi^2$ test to choose the one that better reproduces the absorptions in the data.
We obtained the different models of pure $CH_{4}$ by changing the grain size, and
then we applied shifts to the data to select the combination of shift and grain size that minimized the $\chi^2$.
To evaluate the error associated with the shifts and grains sizes, we repeated the process
several times, randomly changing the upper (longer wavelength) and lower (shorter wavelength) limits of the bands 
within an interval of 10-20 \AA. We then calculated the average and the standard deviation of the shifts and the grain sizes obtained. 
In the case of the shifts, we also considered the errors due to the calibration of the spectra (0.01, 0.04, and 0.02 \AA\ for phase 1, phase 2, and phase 3, respectively).
In Fig. \ref{fit} we show an example of a comparison of a model of pure methane ice with an absorption band of one of our spectra.
The values of the shifts obtained for each absorption band with respect to the theoretical value, 
the size of the grains 
corresponding to the best fit, and the corresponding errors are listed in Table \ref{shift_dim}. 

We were not able to compare all the bands with a model, so some values in Table \ref{shift_dim} are missing.
In the case of the band centered at 0.8442 $\mu m$ of the phase 2, it was not possible to fit a model properly 
because of a bump in the relative reflectance at $\sim$ 0.85 $\mu m$, itself probably caused by an observational artifact that changes the profile of the band. 
For the bands centered at 0.7296, 0.7862, 0.8691, and 0.9019 $\mu m$ of phase 3, 
the S/N is too poor to give a reliable measurement of the center of the bands.

\begin{table*}
\begin{center}

          \begin{tabular}{lcccccccc}
            \hline
            \hline
             \textbf{$CH_{4}$ band} & \multicolumn{2}{c}{\textbf{phase 1}} & & \multicolumn{2}{c}{\textbf{phase 2}}& &\multicolumn{2}{c}{\textbf{phase 3}} \\
             center ($\mu m$) & shift (\AA)& grain size ($\mu m$) & &shift (\AA)& grain size ($\mu m$)& &shift (\AA)& grain size ($\mu m$)\\
             \hline
             \\
             0.7296 & $-3.8_{-1.8}^{+1.5}$ & $4450^{+50}_{-50}$ & & $-2.2_{-3.0}^{+2.5}$  & $3550^{+50}_{-50}$ & & $---$ & $---$ \\
             \\
             0.7862 & $-2.8_{-4.8}^{+3.2}$ & $5850^{+50}_{-50}$ & & $-4.2_{-4.0}^{+3.8}$ & $9800^{+50}_{-50}$ & & $---$ &$---$  \\
             \\
             0.7993 & $-3.0_{-3.3}^{+3.9}$ & $5200^{+100}_{-100}$& &$-4.8_{-5.6}^{+2.0}$  & $10000^{+50}_{-100}$ & & $-2.3_{-3.4}^{+5.2}$ & $7000^{+50}_{-200}$ \\
             \\
             0.8442 & $-3.2_{-4.0}^{+2.0}$ & $3750^{+200}_{-100}$& &$---$&$---$& &$-3.8_{-4.0}^{+3.5}$  & $4100^{+400}_{-100}$ \\
             \\
             0.8691 & $-3.3_{-4.0}^{+3.5}$ & $1750^{+50}_{-50}$ & & $-4.1_{-2.6}^{+3.2}$  &$2000^{+50}_{-50}$  & & $---$ & $---$ \\\
             \\
             0.8897 & $-3.8_{-2.0}^{+2.1}$ & $1300^{+50}_{-50}$ & & $-4.1_{-2.0}^{+1.8}$  &$1450^{+50}_{-50}$  & &$-3.7_{-1.7}^{+3.5}$  &$1150^{+50}_{-50}$  \\
             \\
             0.9019 & $-4.6_{-2.0}^{+3.2}$ & $1200^{+50}_{-50}$ & & $-3.6_{-1.6}^{+5.2}$ &$1550^{+50}_{-50}$  & &$---$  &$---$  \\
             \\
             \hline
          \end{tabular}
          \caption{Shifts of the center of the bands of $CH_4$ in the spectrum of Makemake obtained from the comparison 
          with reflectances of pure methane ice. The grain size associated to the model that is the best fit to the data is also included.}
\label{shift_dim}
   
\end{center}
      
   \end{table*}

\section{Discussion and conclusions} \label{dis}

We observed Makemake during a time interval of 6.3 hours, obtaining three combined spectra that cover $\sim 70\%$ of its rotational period valid for the analysis. 
Each spectrum is formed by two parts, one covering the $0.35$ - $0.52\ \mu m$ range and one covering the $0.70$ - $0.92\ \mu m$ range.

The spectral slope varies from $20 \pm 4$ to $32 \pm 4\%/1000$ \AA\ (see Table \ref{slope}), which is a significant variation that is larger than the uncertainties. 
This could be indicative of a color variation related to differences in the distribution or the particle size of the solid organic compound 
along the surface of Makemake with rotational phase \citep{moroz, brunetto_2006}, 
as observed for Pluto \citep{grundy1996,buratti,Buie_b}.  

It is interesting here to compare the slope of our spectra with the slope of other spectra 
from Makemake or from other methane-ice rich dwarf planets calculated by us in the same way. 
That comparison is shown in Table \ref{slope}. The slope of the visible spectrum of Makemake acquired in 2005 by \cite{licmak}  
is $26 \pm 4\%/1000$ \AA. 
It is similar to the slope of our three phases, but also similar to that of an average Pluto spectrum obtained with the WHT 
\citep[Pinilla-Alonso, private communication,][]{cruik_dps} 
and twice (on average) that of Eris, from the spectrum in \cite{Alvaro}.
Objects with such red surfaces can also be found predominantly in the population of red Centaurs and classic TNOs. 
There are multiple examples in the literature of spectra with similar slopes that are modeled using tholins \citep{noe_quaoar, cruik_1998}. 

It has been assumed for many years that complex-organic solids are present on their surfaces \citep{roush}. 
Therefore, the red slope of the spectra presented in this paper suggests that the surface of Makemake would be better represented 
by the surface of Pluto with a higher amount of organics, which are products of the processing of nitrogen, 
methane, and other volatiles, than by the surface of Eris, which as suggested by \cite{liceris} and \cite{Alvaro}, 
may have been refreshed by the collapse of an atmosphere of volatiles. In Section 3 we gave the shifts of the center of methane-ice absorption bands, and
the size of the grains representing the depth of the layer that produces the bands (see Table  \ref{shift_dim}).

\begin{figure}
   \begin{center}
   \includegraphics[width=9cm]{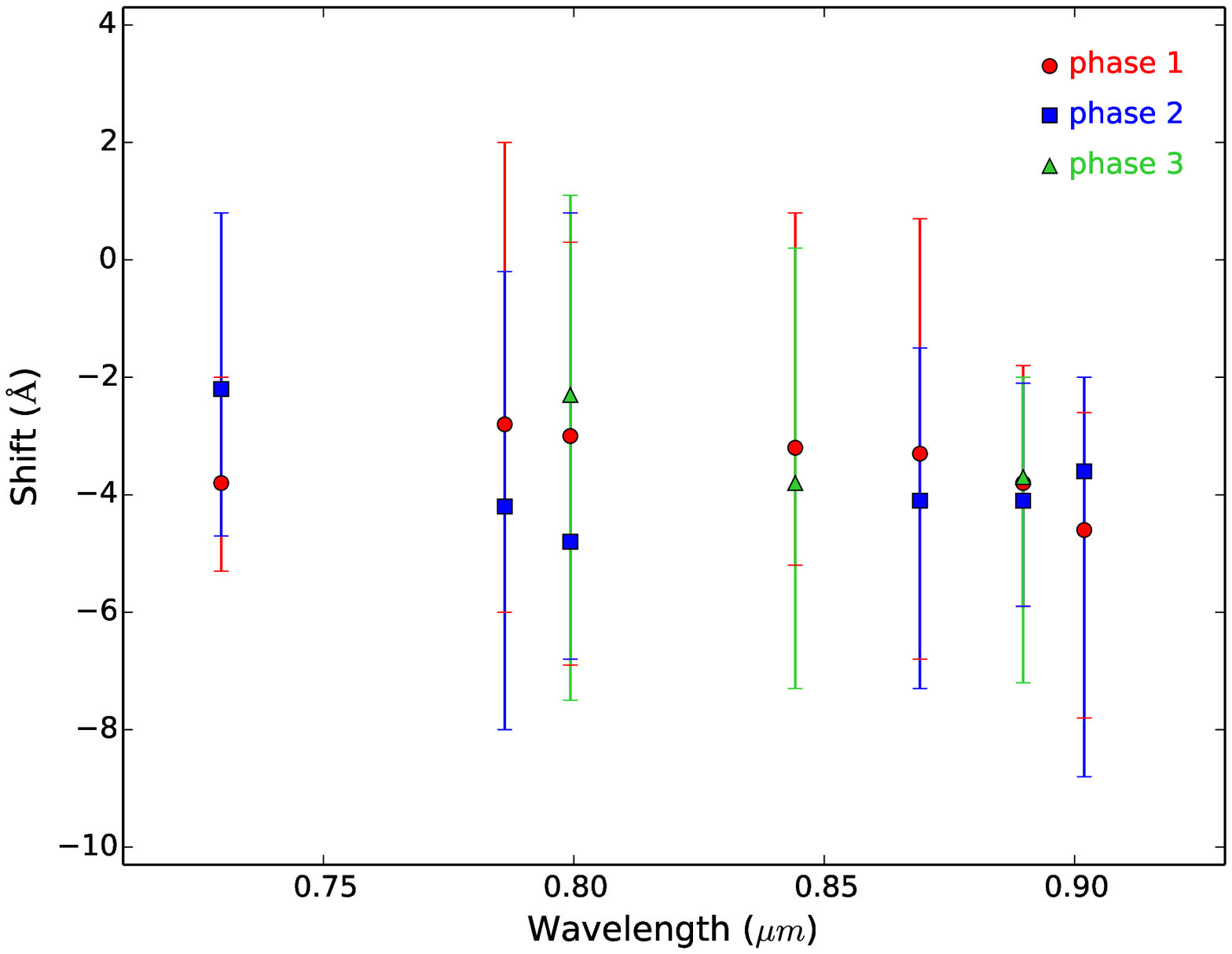}  
   \includegraphics[width=9cm]{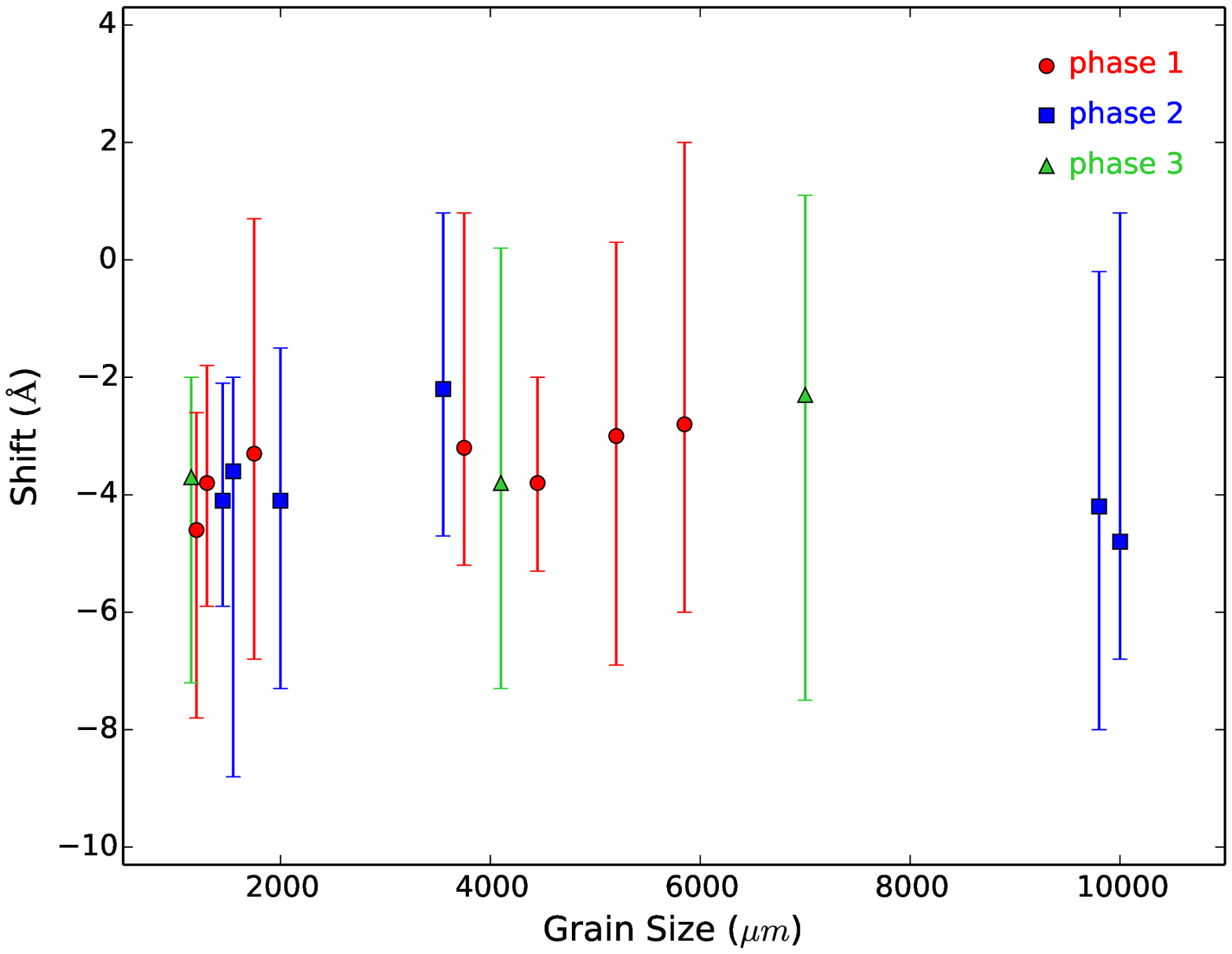} 
   \end{center}
      \caption{Top panel: shifts of the center of each methane-ice band for the three phases. 
      Bottom panel: plot of the shifts of the center of the bands and the corresponding size of grain, given by the model fit, for the three phases. 
      In both panels, the red circles correspond to phase 1, the blue squares to phase 2, and the green triangles to phase 3.}
         \label{sh_lam}
\end{figure}

Previous works \citep{licmak, tegler2007, tegler2008} found that the centers of the absorption bands of methane ice 
in the visible and near-infrared  spectra of Makemake's surface are shifted blueward, indicating that methane is in solid solution 
with nitrogen \citep{quirico}. 
To study possible rotational variations we compared the shifts at the three phases observed for each band (Table \ref{shift_dim} and top panel of Fig.\ref{sh_lam}). 
We do not find any variation within the errors, which suggests a homogeneous distribution of the $CH_4/ N_2$ dilution over the 
71\% of the surface of Makemake sampled by our observations.

\begin{figure}
   \begin{center}
   \includegraphics[width=9.3cm]{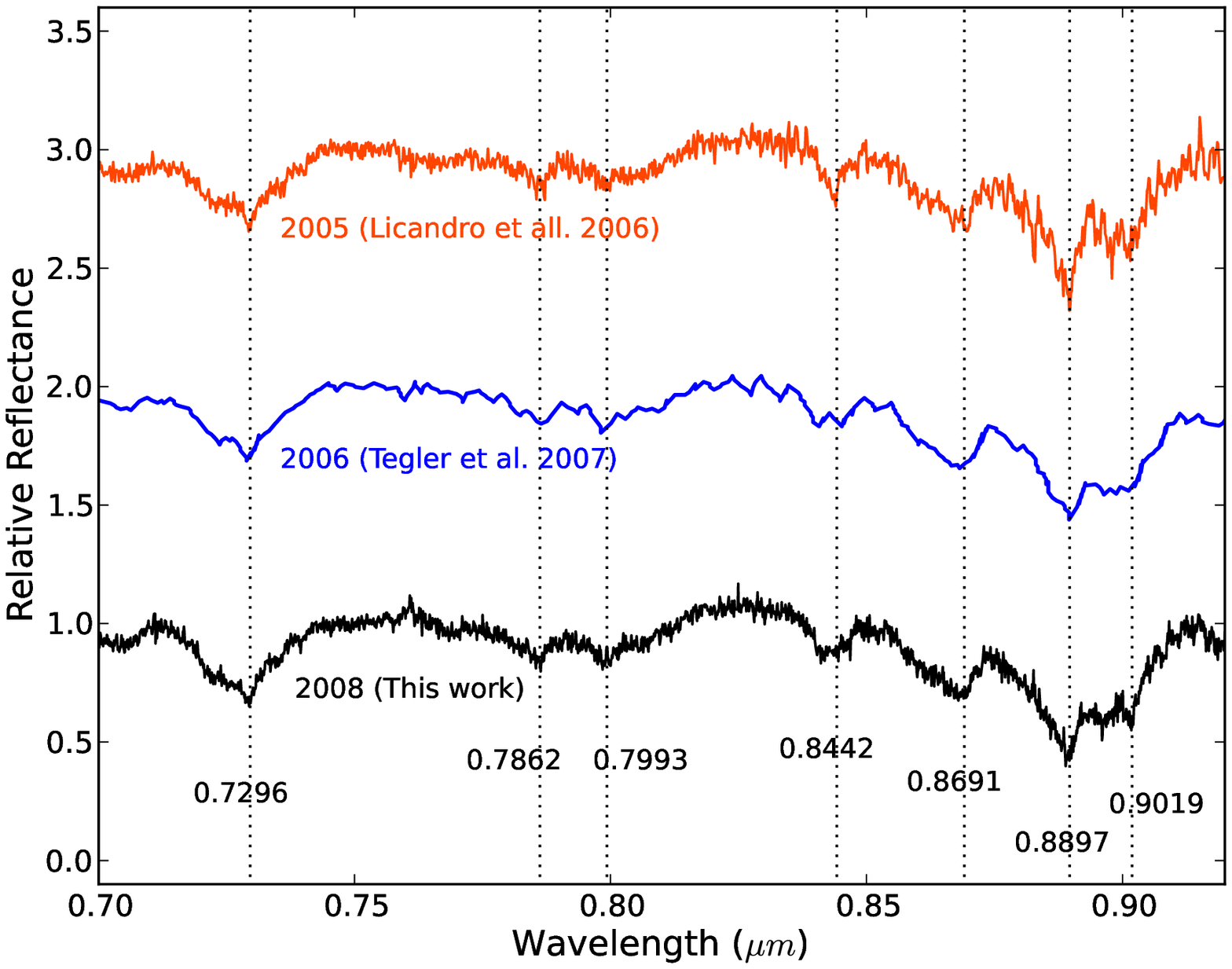}  
   \end{center}
      \caption{Comparison of the spectra of Makemake from the literature in the $0.70$ - $0.92\ \mu m$ range. From top to bottom: in orange, the spectrum from \cite{licmak} taken in 2005; 
      in blue the spectrum from \cite{tegler2007} taken in 2006; in black the spectrum corresponding to phase 1 from this work taken in 2008. 
      All the spectra are normalized at $0.75\ \mu m$ and shifted by 1.0 in relative reflectance for clarity. 
      The dotted lines denote the center of the absorption bands of pure methane ice.}
         \label{red_all}
\end{figure}

All $CH_4$ absorption bands in our spectra show a similar shift of the center toward shorter wavelengths with a mean value of  $-3.6\pm 2.9$ \AA. 
This agree with the shift values reported by \cite{licmak} of $-6$ - $0 \pm 7$ \AA\  
and \cite{tegler2007,tegler2008} of $-4 \pm 1$ \AA. 
In Fig.\ref{red_all} we show, from top to bottom, Makemake's spectrum from \cite{licmak} taken in 2005, the spectrum from \cite{tegler2007} taken in 2006, 
and the spectrum corresponding to first phase reported in this paper, taken in 2008.
The three spectra are very similar, indicating no significant variation in surface composition in the time frame of these three years. 
This is not surprising given the small variation in the heliocentric distance of Makemake in this period, only 0.16 AU, which indicates that the solar insolation 
and the surface temperature have been quasi-constant from the first to the last observation. 

As mentioned, each band carries information about the degree of dilution at a different depth. 
This depth is represented by the grain size used in the model (see Table \ref{shift_dim}). 
To study the vertical distribution of the $CH_4/ N_2$ dilution, we compared 
the measured shifts (related with the $CH_4/ N_2$ dilution) 
and the depth of the layer associated with each band, computed from our models as the grain size given by the best fit (see Fig.\ref{sh_lam}, bottom panel).
Our results do not show any variations in the shifts with the depth within the uncertainties (always greater than 1 \AA),
also taking into account that other factors, apart from $CH_4/ N_2$ dilution, affect the shift of the bands. 
\cite{quirico} measured, for a given dilution level, a difference of only 1\AA \ in the shift for the bands  $ 0.8897\ \mu m$ and $ 0.9019\ \mu m$, 
well below the uncertainties of our measurements. 

We therefore conclude that these results suggest a certain degree of homogeneity in the degree of dilution over the first 10 mm of Makemake's surface. 
Similar results were found for Pluto, Eris, and Triton in \cite{tegler2012}, 
where they claim that the first few centimeters are homogeneous.


The analysis of Makemake spectra allow us to conclude that:

\begin{itemize}

\item The three spectra present a red slope between 0.40 $\mu m$ and 0.52 $\mu m$, which is compatible with the spectral slope reported by \cite{licmak}. 
The slopes of Makemake spectra are similar to that of Pluto and double that of Eris.
This suggests that the surface of Makemake would be best represented by the surface of Pluto, 
with a high amount of solid organics,  
and not the surface of Eris, with a higher amount of volatiles.
\item The spectral slope varies from $20 \pm 4$ to $32 \pm 4\%/1000$ \AA\  along the rotation covered by our data, 
a variation that is significantly larger than the uncertainties. 
This may be attributed to a color variation, which would indicate a difference in the distribution or in the grain size of the solid organic compound over the surface of Makemake. 
However, this needs further confirmation. 
\item In all spectra the $CH_4$ bands are blue-shifted compared to those of pure methane ice by $\sim -4$~\AA\  (a mean shift of $-3.6\ \pm\ 2.9$ \AA) in agreement with previous measurements, 
supporting that methane is in solid solution with nitrogen in the surface of Makemake. 
As noted before, the lower shift of the bands compared with the shifts measured for Pluto and Eris 
indicate less $N_2$ on the surface of Makemake, as predicted by the models of the retention of volatiles for this object. 
\item The shifts measured for each individual band are very similar for the three phases. 
These results suggest that there is no rotational variation, or in the event of this variation being present, it would be very small. 
New observations at higher S/N are needed to make an accurate statement about this topic.
\item The comparison of the measured shifts of all the bands are similar, within the uncertainties, suggesting that there are no large variations 
in the degree of dilution of $CH_4/ N_2$ with the depth along the 10 mm sampled by our observations.
\item Comparison with spectra available in the literature shows no evidence of secular variations for the surface composition of Makemake between 2005 and 2008, 
as suggested by the small changes in solar insolation along this period. 
It would, at any rate, be good to know the rotational period of Makemake with better precision, 
so that spectra taken at different epochs could be cophased.
This would allow a better comparison, which should permit any rotational to secular variation to be separated.

\end{itemize}

\begin{acknowledgements}
We want to thank Dr. William Grundy for his useful comments that helped to improve this manuscript. 
VL and JL gratefully acknowledge support from the Spanish  
``Ministerio de Econom\'ia y Competitividad'' (MINECO) projects  
AYA2012-39115-C03-03 and ESP2013-47816-C4-2-P. 
Based on observations made with the William Herschel Telescope operated on the island of La Palma by the Isaac Newton Group 
in the Spanish Observatorio del Roque de los Muchachos of the Instituto de Astrof\'isica de Canarias. 

\end{acknowledgements}


\bibliographystyle{aa} 
\bibliography{biblio} 

\end{document}